# GPTCloneBench: A comprehensive benchmark of semantic clones and cross-language clones using GPT-3 model and SemanticCloneBench

Ajmain I. Alam*, Palash R. Roy*, Farouq Al-omari, Chanchal K. Roy, Banani Roy, Kevin A. Schneider
Department of Computer Science, University of Saskatchewan, Saskatoon, Canada
E-mail: {ajmain.alam, palash.roy, faa634, chanchal.roy, banani.roy, kevin.schneider}@usask.ca

*Abstract*—With the emergence of Machine Learning, there has been a surge in leveraging its capabilities for problem-solving across various domains. In the code clone realm, the identification of type-4 or semantic clones has emerged as a crucial yet challenging task. Researchers aim to utilize Machine Learning to tackle this challenge, often relying on the BigCloneBench dataset. However, it's worth noting that BigCloneBench, originally not designed for semantic clone detection, presents several limitations that hinder its suitability as a comprehensive training dataset for this specific purpose. Furthermore, CLCDSA dataset suffers from a lack of reusable examples aligning with real-world software systems, rendering it inadequate for cross-language clone detection approaches. In this work, we present a comprehensive semantic clone and cross-language clone benchmark, GPTCloneBench [1] by exploiting SemanticCloneBench and OpenAI's GPT-3 model. In particular, using code fragments from SemanticCloneBench as sample inputs along with appropriate prompt engineering for GPT-3 model, we generate semantic and cross-language clones for these specific fragments and then conduct a combination of extensive manual analysis, tool-assisted filtering, functionality testing and automated validation in building the benchmark. From 79,928 clone pairs of GPT-3 output, we created a benchmark with 37,149 true semantic clone pairs, 19,288 false semantic pairs(Type-1/Type-2), and 20,770 cross-language clones across four languages (Java, C, C#, and Python). Our benchmark is 15-fold larger than SemanticCloneBench, has more functional code examples for software systems and programming language support than CLCDSA, and overcomes BigCloneBench's qualities, quantification, and language variety limitations. GPTCloneBench can be found here[1].

*Index Terms*—Software Clone, SemanticCloneBench, GPT-3, Language Model, Machine Learning, Cross Language Clone, Semantic Clone, BigCloneBench.

## I. INTRODUCTION

Code clones, which refer to identical or nearly identical code snippets in software systems, have long been studied in software engineering. It manifests when programmers use the source code knowledge base already in place to annex new features onto the same or distinct software systems/platforms. According to studies, a software system may have between 9% and 17% of code clones [5], [58]. They can be divided into four categories: Type-1 (exact copies except formatting differences and comments), Type-2 (syntactically similar fragments), Type-3 (near miss clones), and Type-4 (semantic clones), with Type-3 and Type-4 clones being the most complex and challenging to detect.

Because of its advantages during the development process, cloning is a crucial practice for programmers [35]. Finding clones is essential since they harm the quality of the software later in its life cycle and/or could introduce bugs during adaptations [2], [13], [57]. In order to detect all sorts of clones, syntactic [6], [8], [31] and semantic [10], [15], [18], [38], a large number of clone detectors have been developed. For each type of clone, practitioners need to be aware of the clone detection tool's accuracy [6], [32], [43]. Hence, the recall and precision of these detection tools must be assessed.

In order to evaluate the performance of code clone detection tools, benchmarks [1], [6], [46] have been developed. In recent times, ML has been explored and used to detect clones and code similarities other than traditional clone detection tools. While BigCloneBench [46] has primarily been built for evaluating the recall of contemporary clone detection tools, it has also been widely used training innovative machine learning based approaches to Type-4 semantic clone detection [47]. However, BigCloneBench has not been designed for this purpose and has a number of limitations for being used as training dataset for semantic clone detection. Due to the design principles of how the benchmark was created, imbalance issues have been identified, including the ambiguity in the definition of semantic clones [16]. Thus, ML-based clone detection algorithms trained on BigCloneBench may overlook semantic clones or report incorrect results [16], [47]. Furthermore, it has only Java language clones.

Al-omari et al. [1] build SemanticCloneBench that has semantic clone pairs for four different programming languages, such as Python, Java, C, and C#. This benchmark consists of 1000 clone pairs for every language. Its clones have been selected randomly from the crowd contributors in the Stack Overflow community. Unfortunately, while SemanticCloneBench has a diverse and balanced functionality for evaluating existing tools, the number of clones may not be sufficient for ML-based training for semantic clone detection.

---

*Both authors contributed equally.
[1]GPTCloneBench: https://shorturl.at/jvxOV

Yu et al. [54] have proposed an updated version of the BigCloneBench dataset by abstracting the identifier names to enhance its utility. As reported by [54], the BigCloneBench dataset typically contains semantic clones (MT3, WT3/T4) that share identical identifier names. The authors conducted experiments using various machine learning-based clone detectors and highlighted that altering the identifier names hinders the detection of these semantic clones. However, it is noteworthy that the authors of [54] solely focused on updating the identifier names without addressing potential modifications to the implementation while retaining the same functionality.

Finally, CLCDSA [26] is the only dataset available for cross-language clones, according to our knowledge. The problem with this dataset is it only contains cross-language clones of different programming contest solutions, which are considered toy data according to the software clone community, as there is no use in the software industry or simply not reusable in the software industry and by no means they represent real-world clones.

Despite the existence of a number of code clone benchmarks, none of them is big and comprehensive enough for the purpose of machine learning to detect semantic clones. Therefore, in this research, we built a semantic and cross-language clone benchmark, GPTCloneBench using the GPT-3 model focusing on solving these diverse problems of benchmarks for ML-based clone detection tools [16] [47]. Our approach involves utilizing code fragments from SemanticCloneBench and prompting the GPT-3 model to generate semantic code for these specific fragments by means of query formulation. We were able to generate more than 66k clones for both the same language and across languages. We used NiCad to filter out the syntactic clones before we manually validated all the clones. We also validated a random sample of our benchmark by conducting functionality testing for both fragments of a clone pair and making sure they produced the same output for the same input. We also exploited an automated clone valiation tool, CloneCognition [24] [23], an ML clone evaluation tool, to further validate the generated clones. In our benchmark, we have 37,149 true semantic clones, 19,288 syntactic clones (Type-1 or Type-2 clones) and 20,770 cross-language clones. The gap between semantic clones and syntactic clones is low; as a result, this mitigates the problem related to the imbalance and bias as we have an almost equal number of semantic and non-semantic clones. Finally, we have tested a selected clone detection tool's performance on GPTCloneBench. We evaluated SourcererCC [37] and Oreo [36] for semantic clone detection and CLCDSA [26] for cross-language code clones.

For the remaining parts, we have organized our paper as follows. Section II discusses the background of our study, where we have explained the different types of clones, the GPT-3 model, and clone detection tools. The detailed architecture of GPTCloneBench is described in Section III. In Section IV, we gave about the validation procedure for the benchmark, and in Section V, we tested different clone detection tools using our benchmark and analysed the results. In Section VI, we have focused on describing the related work. In Section VII, we have talked about the threats to the validity of our research, and in Section VIII, the conclusion of our research is discussed.

## II. BACKGROUND

### A. *Code Clone*

When one piece of code in the source code is identical or nearly identical to another piece of code in the code base, we refer to the first as a *code clone* to the second, and we call the two together a *clone pair*. The idea of resemblance serves as the foundation for this definition. Following is a classification of the definition of clone that has found widespread acceptance in the scientific literature. [34]

- **Type-1 (T1)**: Code segments that are identical except whitespace differences (and sometimes layout differences) and comment differences [6].
- **Type-2 (T2)**: Fragments that are structurally and syntactically identical to one another, with the exception of differences in identifiers, literals, types, layout, and comments. [6]
- **Type-3 (T3)**: Fragments that were copied with additional alterations made. Changes can also be made to literals, types, layouts, and comments, in addition to identifiers, which can be renamed or removed entirely. Statements can be modified, added to, or removed entirely [44].
- **Type-4 (T4)**: Two or more snippets of code that, when combined, carry out the same computation but do so in accordance with distinct syntactic variations [44].

The classification of code clones into Type 1 and Type 2 is based on their textual similarity within the context of code clones. Code fragments are considered clones if they display textual similarity, regardless of whether their functionality differs. Textual clones are a commonly observed form of clones in software codebases, which frequently arise due to the practice of copying/pasting.

In contrast, the identification of semantic clones poses a greater challenge due to their potential to be executed through diverse syntactical structures. Textual clones are characterised by identical text, whereas semantic clones exhibit similar functionality but may have been implemented using distinct syntactic structures. The identification of semantic duplicates often necessitates the utilisation of advanced methodologies that scrutinise the operational characteristics of code segments.

Various definitions for semantic clones have been proposed in the academic literature concerning code clones.

The scope of definitions encompasses a variety of terms, including but not limited to relative clones [39], redundant code [19] [41], dependent clones [10], functional clones [12], functionally similar clones [40] [14] [48] and Type 4 clones [35]. Although researchers generally agree that semantic clones exhibit comparable functionality but are implemented using distinct syntax, there remains a lack of agreement regarding the particular form of semantic similarity that defines these clones.

There exists a divergence among researchers in the definition of semantic clones, with some adhering to a limited interpretation that pertains to a particular form of semantic resemblance and others embracing a more comprehensive and less precise construal of the concept. Irrespective of the particular definition employed, there is a general consensus that semantic clones refer to clones possessing identical functionality that is executed using distinct syntax [4] [6].

In our study, we have followed the categorization scheme of BigCloneBench [42], [44] for code clones based on their similarity percentage with slight considerations for a grey area. Moderately type-3 (MT3) [44] clones have been defined as those with a similarity of 50%-70%, with an additional 5% grey area. Clones with a similarity of 50%-75% are classified as Type-3 (T3) clones. Weak Type-3(WT3) [44] clones are considered type-4 clones with a similarity percentage of 0%-50%.

Furthermore, Svajlenko et al. [42], [44] have described that if two code snippets conduct the same computation but they have implemented in different syntactic ways, then they can be considered as Type-4 clones. Based on that, cross-language code clones can be considered to be semantic clones because they share the same functionality and logic despite being written in different programming languages. We have identified cross-language clones as Type-4 clones. This finding indicates that not all semantic clones are cross-language clones, but all cross-language clones fall under this category.

Our categorization scheme is intended to aid researchers in identifying and analyzing different types of code clones. It is important to note that other researchers may use different definitions for the various types of clones, and thus the adoption of a common taxonomy would be beneficial to facilitate better comparison and consolidation of research findings in this area.

Clones of Type 1, Type 2, and Type 3 are classified according to their level of textual similarity. Clones are identified among code fragments when they exhibit textual similarity, regardless of their functional differences. Textual duplication is a prevalent occurrence in software codebases, often resulting from the practice of copying and pasting code. In contrast, identifying semantic clones poses a greater challenge as they may be executed through varying syntactic means. Previously we have put forth the definitions of semantic clones and discrepancies among scholars. Nonetheless, all definitions cited concur that semantic clones possess identical functionality and are executed through distinct syntax.

### B. *Generative Pretrained Transformer-3 (GPT-3)*

Brown et al. [7] have introduced the GPT-3 model, which is a successor of GPT-2 model [29]. Brown et al. [7] have trained 8 different models where the parameters range from 125 million parameters to 175 billion. For our research, we have focused on the "Text-Davinci-003" model. Text-Davinci-003 has gained a lot of popularity because of its robustness and its ability to produce long-form content [22], [11], [21]. Text-Davinci-003 model consists of 96 layers, 12288 units in each bottleneck layer, and each attention head's dimension is 128. The Text-Davinici-003 model is trained on 3.2 million batch-size data and a 0.00006 learning rate.

### C. **Clone Detection Tools**

*1)* **NiCad:** Roy and Cordy. [31] presents a software tool for detecting code clones in software systems. The tool is called NiCad (for "Near-miss Clone Detector"), and it is capable of detecting different types of clones, including Type-1, Type-2, and Type-3 clones. The NiCad clone detection approach uses a combination of text-based and tree-based approaches with support for different flexible source code normalization and transformation features. The tool provides a range of options for configuring the detection process, including parameters for defining the minimum size of clones and the level of similarity required to consider two code fragments as clones.

*2)* **SourcererCC:** Sajnani et al. [37] propose a scalable and effective approach for detecting code clones in large codebases, which can aid in software maintenance, bug fixing, and software evolution. It leverages a combination of tokenization, hashing, and indexing techniques to identify code clones across a codebase efficiently. SourcererCC is designed to work with various programming languages, making it a versatile tool for detecting code clones in diverse codebases.

*3)* **CLCDSA:** Nafi et al. [26] propose an approach to detect cross-language code clones, where the code fragments are written in different programming languages. The proposed approach, called CLCDSA, leverages a combination of syntactical features and API documentation to detect cross-language code clones. CLCDSA uses a code representation technique that captures the syntactical structure of code fragments and a technique that incorporates API documentation to compare the functionality of code fragments. The approach was evaluated on a dataset of code fragments written in different programming languages, including C++, Java, and Python. It demonstrated high precision and recall rates compared to state-of-the-art cross-language clone detection tools.

*4)* **Clone Cognition:** Mostaeen et al. [24] present a machine learning-based approach for validating code clones detected by other clone detection tools. The proposed approach, called CloneCognition, uses machine learning techniques to classify detected clones as either true clones or false positives. CloneCognition works by extracting a set of features from detected code clones, such as the similarity of syntax, the similarity of function names, and the number of lines of code. These features are then used to train a machine-learning model, which can classify new code clones as true clones or false positives.

*5) Oreo:* The Siamese Network architecture is utilised by Oreo [36] for the purpose of predicting code clones. In this approach, Java code fragments are represented by software metrics. The utilisation of a blocking technique that involves filtering based on comparable size and satisfactory overlap of "action tokens" is implemented. The provision of the training supervision signal can be facilitated by employing a state-of-the-art code clone detector, such as SourcererCC [37]. The study conducted by the authors demonstrates that the model produced exhibits exceptional scalability, competitive performance, and remarkable proficiency in the "twilight zone" of code clones, characterised by moderately Type-3 and beyond. This tool is trained on 50k GitHub Java projects.

## III. Building GPTCloneBench

The process of generating GPTCloneBench is demonstrated in Figure 1. It starts with extracting code functions from SemanticCloneBench. To facilitate this, an automated script was devised to detect potential functions that met the desired functionality. Once identified, a query was formulated, employing the function from the previous stage to extract a response from the GPT-3 language model. The output was then processed using an automated script to generate input and corresponding output files based on the input log. To ensure that the resulting clone pairs were of high quality, we eliminated textual similar pairs (input vs output, whose similarity is more than 75%) by using NiCad. Following this, manual validation was conducted to further refine the results by removing any additional clone pairs that did not meet the established criteria. Lastly, the benchmark is finalized by retaining only those clones that passed the aforementioned evaluation criteria.

### A. Select Target Functionality from Semantic-CloneBench

The process of generating a semantic clone benchmark starts with selecting the first clone fragment in the clone pair. To have a diverse functionality that does not belong to a certain programming domain or developer background, we selected functions from SemanticCloneBench. SemanticClonebench extracted its clones from Stack Overflow answers that are submitted by versatile contributors in the programming community, which cover a variety of programming problems, such as file I/O, string manipulation, sorting, and data structures.

### B. Give prompt to GPT-3: Query Formulation

A well-defined prompt is a crucial component of natural language processing tasks. A prompt combines instruction, context, input data and output indicator. Where the instruction is a clear and concise statement of the task or goal for the model, context is any relevant information or background knowledge that can help the model understand the task or domain; input data is what the model needs to process or analyze to produce the output. The choice of prompt technique can significantly impact model performance. There are several techniques to do the prompts, such as Zero-shot [28], [52], Few-shot [9], [52], Chain of Thought [52], Self Consistency [51], and Generate Knowledge [53] [27]. Zero-shot uses a single prompt to perform multiple tasks without any training data or examples, Few-shot uses a few examples of input-output pairs to guide the model to perform a specific task, and Chain of thought uses multiple prompts in sequence to enable the model to reason and learn from its own outputs, Self Consistency uses prompts that check or verify the model's output against its input or other sources of information and lastly Generate knowledge uses prompts that ask the model to generate new facts or information based on its inputs or context. Each technique can be used based on the specific task, data, and model architecture. Selecting the appropriate prompt technique can significantly improve model performance and accuracy. In our case, we used the Few-shot prompting technique to receive our data. For our prompt, we had textual instructions and an input sample to guide the GPT-3 model on what type of output we were expecting.

Numerous queries were experimented with to elicit a response from the GPT-3 language model. We have tried with the queries as of Figure 2.

We started by analysing a sample of size 20 for each query. First, we manually evaluate the results and tag them as true semantic clones or false semantic clones(Type-1/Type-2). Finding that GPT-3 is able to create true clones for all the submitted methods. However, the generated clones are of different types. Since we are targeting semantic clones only we measure the syntactic similarity for clones generated by each query. Table I shows the average textual similarity percentages of clone pairs for every query. Interestingly, the various queries exhibited different behaviours for each programming language. However, the prompt "Give me 10 distinctive implementations for the following code <code fragment>" proved to be the most effective, consistently yielding real semantic (non-syntactic) clones in all the languages tested. As a result, we have run it 2 times to facilitate the benchmark.

A customized tool (depicted in Figure 3) was developed to automate the input retrieval from the SemanticCloneBench, prompt generation using the API, and log the resultant output. Initially, a script was designed to extract a code segment from the SemanticCloneBench file.

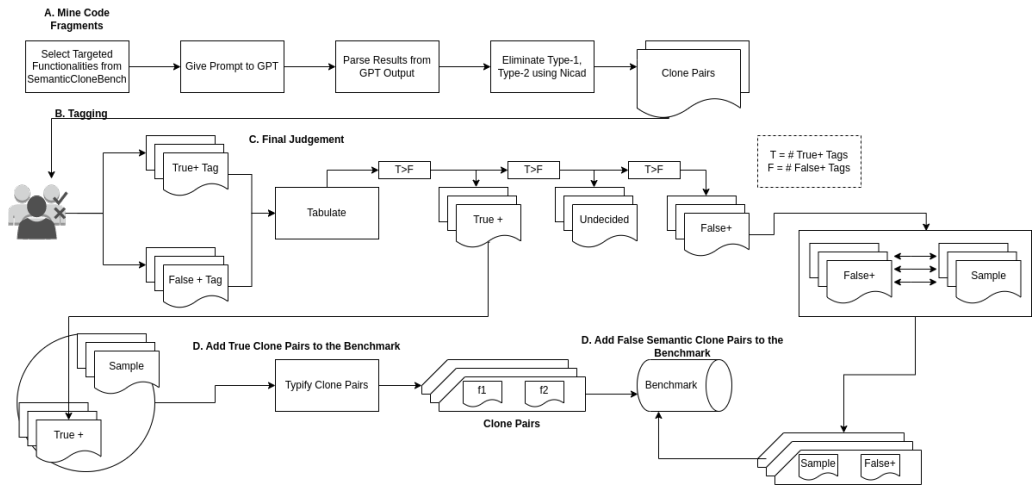

Fig. 1. GPTCloneBench complete process

- Give me 4 **Type-3, Type-4 clone** for the following code *<code fragment>*
- Give me 10 **distinctive** implementation for the following code *<code fragment>*
- Give me **different** implementation for the following code *<code fragment>*
- Give me **unique** implementation for the following code *<code fragment>*
- Give me different **functional** implementation for the following code *<code fragment>*
- Give me different **syntax** implementation for the following code *<code fragment>*

Fig. 2. Query prompts to GPT-3

The code fragment was then passed on to the GPT-3 prompt, where a prompt was formulated based on the query type and conveyed to the model through the API. Subsequently, GPT-3 processed the prompt and generated the corresponding output. To facilitate further processing, the generated output was logged in a file for subsequent analysis and refinement. This automated tool served as a critical component in generating the benchmark dataset with diverse prompt types for evaluating the language model's ability to generate distinctive clone implementations.

TABLE I
Summary Results of similarity % for each query

| Language | Type-3/4 | Different | Unique | Functional | Syntax | Distincitive |
|---|---|---|---|---|---|---|
| Python | 74.75 | 40.4 | 36.5 | 27.1 | 43.1 | 31.9 |
| Java | 100 | 34.55 | 34.55 | 59.5 | 66.2 | 34.55 |
| C | 100 | 74.9 | 82.2 | 41.6 | 91 | 46.5 |
| C# | 100 | 67.33 | 63.33 | 81.2 | 100 | 80.5 |

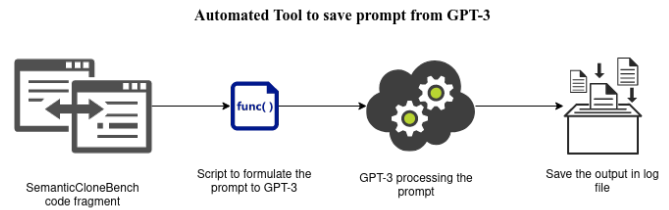

Fig. 3. Prompt automated tool for GPT-3

- Prompt 1: Give me **Type-3, Type-4 clone** for the following code *<code fragment>*
- Prompt 2: Give me 10 **distinctive** implementation for the following code *<code fragment>*

Fig. 4. Query prompts to GPT-3 for Type-3 and Type-4 clones

As a result, we have focused on two prompts to generate Type-3 and Type-4 clones as of Figure 4.

To generate cross-language clones, we have utilized two programming languages, which are Java and C#. The reasons behind choosing only these two programming languages are that they are more object-oriented, and we want to create a functional benchmark. At first, from SemanticCloneBench, we select a Java code fragment and gave GPT-3 the following prompt, "Give me Python, C, and C# implementation of the following code: *<code>*". By running this query, we were able to get 3 different programming language implementations of a given Java code. To further increase the data, we have moved our focus to C# programming language. In the next prompt, we selected a C# code fragment and gave GPT-3 the following prompt, "Give me Python, C, and Java implementation of the following code: *<code>*". By running this query, we were able to get three different programming language implementations of a given C# code.

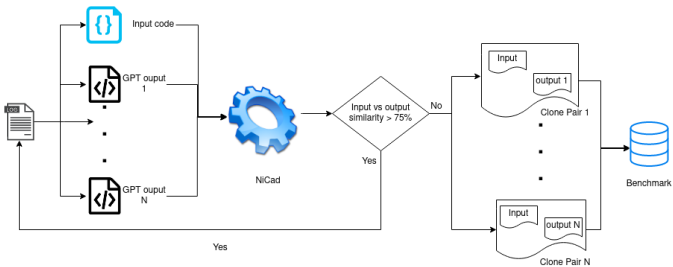

Fig. 5. Automated tool for processing the semantic output of all queries from GPT-3

### C. Parse results from GPT-3

For each prompt, we have followed the same parsing method that we described in this section. In order to validate the output generated by GPT-3 resulting clones, we have performed a two-step evaluation. First, we filtered out syntactic clones by measuring the textual similarity. Second, we manually validate all candidate Type-3, Type-4 and cross-language clones.

*1) **Filtering out syntactic clones**:* In this section, we present our approach to process Type-3 and Type-4 outputs, as illustrated in Figure 5. GPT-3 engine generates a different code for running the same prompt multiple times. We have found that in very few cases, that GPT-3 generates textual output that explains how to implement the input code in a different form without producing any actual code. Also, some of the code produced by GPT-3 may contain syntactical errors or bugs, which could impede the successful execution of the code. To overcome such rare unwanted results and have the best possible semantic clone pair generated by GPT-3, we have run the prompt four times to generate four possible clone pairs to the original method.

At first, from SemanticCloneBench [1], we have taken an input function to create an input file and a separate output file for each GPT-3 output. For instance, in the case of the first query, one input function resulted in four functions generated by GPT-3. Therefore, we created five files for that query: one input file and four output files, which are represented by dot (.) in Figure 5. We then utilize NiCad to identify Type-1 and Type-2 code clones by passing the input and output files through the software. For the NiCad configuration, we selected the threshold of dissimilarity percentage at 99% so that in the metadata file, we can get the similarity percentage, as NiCad cannot detect Type-4 clones. The minimum size for lines in a file is set to 3 as we have some functions that have fewer line numbers. Lastly, we used blind renaming so that it can transform all the identifiers as ID so that we can check the similarity percentage. We compare the similarity between each pair of input and output files and discard any pairs where the similarity exceeds 75%. Conversely, if the similarity between a pair is less than equal to 75%, we save that pair in a file to the benchmark for further validation. Because we have multiple outputs from GPT-3, that is why we can see multiple clone pair files in Figure 5.

In the realm of cross-language clone detection, as the clone pairs are already in different programming languages, there is no syntactical similarity among those. So, using NiCad is infeasible since there is no such syntactic clone in cross-languages. However, generated clones are submitted to the next step for further validation. Furthermore, another validation process (input-output testing) has been adopted to ensure the functional equivalence of code clones, which will be described later.

*2) **Tag Functions**:* After filtering out undesired clones as described in Section III-C1, we have engaged in a rigorous manual validation process. This involved a thorough examination of all code fragments in determining whether filtered data was accurate or not and whether the clone pairs produced the same output for the same input. After file generation, we manually validated the clone pairs to ensure their validity. To facilitate accurate assessment, BigCloneBench's GUI-based Clone Validator [2] was utilized, which provided syntax highlighting for the candidate code and displayed the exemplar functions and specifications alongside the candidate for reference.

In the validation process, a total of nine judges participated, comprising six undergraduate research students and three post-docs. The undergrad students were divided into three groups, each consisting of two students. The dataset was then split into three sections, containing 12758, 12758, and 12760 pairs, respectively, for semantic clones and 7454, 7454 and 7456, respectively, for cross-language. Each group independently assessed their assigned section, labelling the clone pairs as true positive, false positive, or undecided based on their understanding of functionality. For a clone pair to be considered a true semantic pair, both members of the group had to tag it as true. Conflicting results within a group led to excluding that pair from the true pairs listing. All six judges followed this procedure.

During the assessment process, each group in total tagged 951 undecided pairs, which were later shuffled among the three groups. The groups attempted to resolve the undecided status of these pairs by tagging them as either true or false positive and remained with 357 total undecided pairs. Finally, these remaining 357 undecided pairs were collectively assessed and labelled by the three post-docs through discussion. Approximately 510 hours were spent by nine judges to validate. We want to mention that the undergraduate research students were trained and given instructions on the functionalities of why and how we defined the semantic clones.

```
int main () {
    char sentence [] = "this is a sentence";
    char *newSentence = (char *) malloc (strlen (sentence) +1);
    int i, j, start, k;
    start = 0;
    for (i = 0;; i++) {
        if (sentence[i] == ' ' || sentence[i] == '\0') {
            char *word = (char *) malloc ((i - start) + 1);
            for (j = i - 1, k = 0; j >= start; j--, k++) {
                word[k] = sentence[j];
            }
            word[k++] = ' ';
            word[k] = '\0';
            strcat (newSentence, word);
            start = i + 1;
        }
        if (sentence[i] == '\0')
            break;
    }
    printf ("%s\n", newSentence);
    return 0;
}
```
Input

```
int main () {
    char sentence[] = "this is a sentence";
    char* newSentence = malloc (strlen (sentence) + 1);
    int start = 0;
    for (int i = 0; sentence[i]; i++) {
        if (sentence[i] == ' ') {
            for (int j = i - 1; j >= start; j--) {
                strncat(newSentence, &sentence[j], 1);
            }
            strncat(newSentence, " ", 1);
            start = i + 1;
        }
        else if (sentence[i + 1] == '\0') {
            for (int j = i; j >= start; j--) {
                strncat(newSentence, &sentence[j], 1);
            }
            strncat(newSentence, "\0", 1);
        }
    }
    printf("%s\n", newSentence);
    return 0;
}
```
GPT-3 output

Fig. 6. Semantic code clone generation sample

### D. *True Clone Pairs*

In summary, we were able to generate over 79,928 Clone pairs of Type-3, Type-4 and cross-language clones from two prompts. Although the expected number was 110,604 due to GPT-3 token limitation, buggy samples from GPT-3 and some technical errors during the experiment, we generated a total of 57,564 semantic clone pairs and 22,364 cross-language clone pairs. After a thorough screening process and manual validation, we chose 37,149 true Type-3 and Type-4 clone pairs from four different languages: Java, C, C#, and Python. Also, we have a true 20,770 cross-language clone pairs. As mentioned earlier, we have followed the categorization scheme of BigCloneBench/BigCloneEval [44] for code clones based on their similarity percentage with slight considerations for a grey area [36]. In our benchmark, we have defined Moderately Type-3 clones [44] (with a similarity of 50%-75% with the 5% added for the clones in the grey area [36]) as Type-3 clones and Weak Type-3/Type-4 clones [44] (with a similarity percentage of 0%-50%) as Type-4 clones. The reason for considering 5% grey area in Type-3 clones was that we wanted to have more semantic clones in our dataset that were in the twilight zone [36]. Figure 6 shows an example of the generated semantic clones. Figure 7 shows a cross-language clone class for four programming languages. GPT-3 was fed with the Java method, which is given in Figure 7 and generated the code fragments in Python, C and C# based on that. It is clear that the clone is not syntactical and would be classified into Large-gap Type-3 or Type-4 clones. To generate false semantic clones(Type-1/Type-2) from two prompts, we selected those pairs for which NiCad gave textual similarity of more than 75%. The total number of false semantic clone pairs is 19,288.

[2]https://github.com/jeffsvajlenko/ValidateClones

We added the false semantic clones with proper labels so that ML models could extract the semantic and non-semantic clones' features. Because in order to train an ML model to detect semantic clones, we must provide both semantic and non-semantic examples to the machine. It enables the model to learn the patterns and characteristics of both semantic and non-semantic cases. It will assist the model in developing a more accurate knowledge of the problem at hand and making better predictions. Lastly, we designed our GPT Clone Benchmark (GPTCloneBench) into two forms: stand-alone clones and injected clones.

*1) Injecting in a system:* Our goal in developing GPTCloneBench was to create a versatile and reusable system that incorporates various programming languages. To evaluate its effectiveness in comparison to real-world systems, we selected four medium-sized systems identified in the SemanticCloneBench [1] (Table II). To inject Type-3 and Type-4 code clones into different files, we randomly selected locations while ensuring that no clone pairs were injected in the same file. We injected these clone pairs into these real-world projects so that practitioners can test their semantic or cross-language clone detection tools in these datasets.

*2) Stand alone Clones:* We kept all the clone pairs in a single text file for different uses. Inside the stand-alone clone folder, there are two folders named false semantic(Type-1/Type-2) clones and true semantic clones. Practitioners could use a subset of clones for other testing or could inject them into other systems. Some clone detectors cannot scale for large systems and others have certain limitations. Also, if anyone wants to train their machine learning model, they will find it easy to do it.

### IV. VALIDATING GPTCLONEBENCH

GPTCloneBench has primarily been built using manual validation. Results of clone detection tools are usually

```
// Java                                    // Python                              // C                                          // C#
public static boolean longerTF (boolean [] guess) {   def longerTF(guess):                  bool longerTF (bool guess[], int size) {    public static bool LongerTF(bool[] guess)
    int variableTrue = 0;                      variableTrue = 0                       int variableTrue = 0;                    {
    int variableFalse = 0;                     variableFalse = 0                      int variableFalse = 0;                       int variableTrue = 0;
    for (int x = 0;                            for x in range(len(guess)):            for (int x = 0; x < size; x++) {             int variableFalse = 0;
    x < guess.length; x ++) {                      if (guess[x]):                         if (guess[x]) {                          for (int x = 0; x < guess.Length; x ++)
        if (guess [x]) {                               variableTrue += 1                      variableTrue++;                      {
            variableTrue ++;                       else:                                  } else {                                     if (guess[x]){
        } else {                                       variableFalse += 1                     variableFalse++;                             variableTrue ++;
            variableFalse ++;                  return variableTrue >= variableFalse       }                                        }
        }                                                                              }                                            else{
    }                                                                                  return variableTrue >= variableFalse;            variableFalse ++;
    return variableTrue >= variableFalse;                                          }                                                }
}                                                                                                                               }
                                                                                                                                return variableTrue >= variableFalse;
                                                                                                                            }
```

Fig. 7. Cross-Language code clone generation sample

TABLE II
Systems details where clones are injected

| System | Language | Number of files | Line of code |
|---|---|---|---|
| JHotDraw7 | Java | 711 | 130k |
| PostgreSQL-12.0 | C | 1343 | 1368k |
| Mono1.1.4 | C# | 9822 | 5518k |
| django | Python | 2031 | 240k |

evaluated manually to measure precision or by using benchmarks to measure recall. However, in this research, we have built specialized Type-3 and Type-4 benchmarks that we evaluate manually. It will be impossible to evaluate our benchmark using another clone benchmark since other benchmarks contain different clones and are specialized for different types of clones. So, to further check our benchmark, we perform two types of evaluations. First, we perform a functionality test on clone pairs in the GPTCloneBench to make sure that the code fragments in our benchmark are real and run correctly for proper input. Second, we used CloneCognition [24] [23], an ML clone evaluation tool, to make sure that GPTCloneBench contained Weak Type-3 and Semantic clones only.

*1) Clone pair validation with Functionality testing:* Each GPTCloneBench clone pair contains a pair of methods; a clone method is selected from the Stack Overflow answer, which is considered a programmer-developed method, and a GPT-3 (machine) generated method. Even though all clone pairs are manually validated, we performed this test to make sure that GPT-3 generated methods are syntactical and semantically correct. Therefore, we performed unit testing for clone pairs (both methods) to make sure they are executable and give the same results. For unit testing, the first two authors of the paper randomly selected 500 clone pairs each. They manually checked all the functionalities according to the best of their programming knowledge. It took them almost 192 hours to check in a total of 1,000 clone pairs. They mainly checked if, for the given input, both of the clone pairs produce the same output or not. They observed 99% of them functioning accurately. In future, we are planning to utilize unit testing tools to automate this process.

*2) Clone pair validation with CloneCognition:* To have further confidence in our benchmark, we exploited CloneCognition. CloneCognition can validate and classify syntactic clones with an accuracy of up to 87.4%. Therefore, we do not expect it to recognize clones in our benchmark, as we only focused on semantic clone pairs for testing purposes. We used CloneCognition to classify clones in our benchmark. We used this tool in a reverse way, as it is mostly built for syntactic clones. So, the lower the accuracy is, the better the result of our benchmark is because it proves the benchmark is dominated by semantic clones. Results are presented in Table III for the first query (asking for Type-3 and Type-4 implementation), it is evident that CloneCognition falls short in detecting all true clone pairs, with an accuracy of only 0.15 for Python. However, in the case of Java, C, and C# programming languages, CloneCognition's accuracy improved slightly to around 0.37. Nonetheless, it still could not identify all the clones. In Table IV, we can see that CloneCognition's accuracy has dropped significantly compared to the first query. In the second query, we asked for the distinctive implementation of the input code. The highest accuracy occurred for C programming language with 0.18 indicating that our analysis is true.

TABLE III
CloneCognition Results with Type-3, Type-4 Prompt

| Language | Threshold | # Clones | Predicted | Accuracy |
|---|---|---|---|---|
| Python | 76% | 1495 | 229 | 0.15 |
| Java | 76% | 1087 | 397 | 0.37 |
| C | 76% | 563 | 206 | 0.37 |
| C# | 76% | 1033 | 330 | 0.32 |

TABLE IV
CloneCognition Results with Distinctive Prompt

| Language | Threshold | # Clones | Predicted | Accuracy |
|---|---|---|---|---|
| Python | 76% | 4968 | 228 | 0.05 |
| Java | 76% | 4635 | 516 | 0.11 |
| C | 76% | 2853 | 511 | 0.18 |
| C# | 76% | 4025 | 394 | 0.098 |

## V. GPTCloneBench in use

In this section, we evaluated different syntactic, semantic and cross-language clone detection tools with our benchmark. As a testing metric, we have used Precision and Recall to measure the performance of the code clone detection tools.

$$Precision = \frac{TruePositive}{TruePositive + FalsePositive}$$

$$Recall = \frac{TruePositive}{TruePositive + FalseNegative}$$

For our evaluation, we used GPTCloneBench on SourcererCC [37], Oreo [36] and CLCDSA [26]. We calculated these metrics in two ways: one with stand-alone clones (Section III-D2) and the other with the injected system (Section III-D1).

The Table V presents the result of the tools applied to stand-alone clones (Section III-D2). In our benchmark, the primary emphasis is on calculating the recall, as the clones provided to the tools are true semantic clones. Precision, on the other hand, is not a relevant measure for stand-alone clones due to the inherent nature of the clones where all the clones are true positive within the dataset provided to the tools.

### A. Tools Performance

*1) SourcererCC:* We evaluate the performance of SourcererCC with our benchmark (Table V). Result shows that SourcererCC achieved a low recall rate. The result suggests that SourcererCC effectively identifies Type-3 code clones, which aligns well with our benchmark comprising Type-3 data. However, because our benchmark consists of Type-4 clones, the tool's recall is dropped.

*2) Oreo:* To check the performance of ML-based semantic clone detectors on our dataset, we used Oreo [36] with default threshold for both of the queries' filtered results. The results of our evaluation are presented in Table V. Oreo outperformed SourcererCC with an average recall of 0.49 for the stand-alone clones. However, it is noteworthy that while Oreo is a semantic clone detector tool and, therefore, expected to yield better results than SourcererCC, but its performance on our dataset was not as high as anticipated. However, its high performance indicates that our benchmark has more semantic clones.

*3) CLCDSA:* In order to assess the performance of CLCDSA in cross-language clone detection, we first manually evaluated the clone pairs and then executed CLCDSA [26]. At first, we tried to check model performance on stand-alone cross-language clones. From Table V, we can see the performance of that data. From Table V, we can confirm that the CLCDSA model has failed to detect most of the true clone pairs, and the best performance of CLCDSA is reported in Table V for our dataset. We have expected this result as the CLCDSA model is trained on a dataset primarily composed of competitive programming languages, often lacking practical functionality or reusability. In contrast, GPTCloneBench encompasses more pragmatic and reusable code fragments, enhancing its suitability for broader applications. We were unable to produce results for C programming language as CLCDSA lacks that support.

TABLE V
SourcererCC, Oreo and CLCDSA Recall results

| Tool | Language | Granularity | Recall |
|---|---|---|---|
| SourcererCC | Java | Method | 0.135 |
|  | C | Method | 0.12 |
|  | C# | Method | 0.10 |
| Oreo | Java | Method | 0.49 |
| CLCDSA | Java and C# | Method | 0.19 |
|  | C# and Python | Method | 0 |
|  | Java and Python | Method | 0.001 |

In addition to stand-alone clones' performance, we also wanted to analyse how the tools will perform in system-injected environments.

*4) SourcererCC with system injected clones:* We used the system-injected part of our benchmark to evaluate SourcererCC performance and calculated the recall and precision. The results are given in Table VI. Through manual inspection, among those randomly selected 400 clones, 328 clones are valid clones. Among 176 GPTCloneBench clones, the tool was able to detect 58 clones. This demonstrates that even after achieving high precision, the tool exhibited low recall, indicating its failure to detect our benchmark data within the system. In summary, as our benchmark includes Type-4 clones, the tool's ability to recall is low.

*5) CLCDSA with injected clones:* As CLCDSA [26] doesn't have any support to detect clones from a system like SourcererCC, in this experiment, we injected 20 false clones with the stand-alone clones and measured the precision and recall for this scenario. From Table VI, we can see that the CLCDSA model is not performing well in detecting the cross-language clones. Even though it has achieved 0.47 precision, but the low recall indicates that CLCDSA fails to detect the clones properly.

TABLE VI
SourcererCC and CLCDSA results with injected systems

| Tool | Language | Granularity | Precision | Recall |
|---|---|---|---|---|
| SourcererCC | Java | Method | 0.82 | 0.33 |
| CLCDSA | Java, C#, Python | Method | 0.47 | 0.10 |

## VI. Related Work

Regarding the code clone benchmark, there are several existing benchmarks. Bellon's Benchmark [6] stands as a prominent clone benchmark with extensions [25], [43], stemming from Bellon et al.'s 2002 experiment evaluating six clone detection tools against eight software systems [6]. While it measured recall and precision, concerns about accuracy arose due to the benchmark's construction using participating tools' union results, offering relative measurements but not ensuring comprehensive detection [3]. Manual validation complexities are evident, exemplified by Bellon's 77-hour validation of a mere 2% of candidate clones [35]. Even for small systems like Cook, exhaustive manual validation becomes unfeasible, introducing potential human error. Moreover, the unaddressed reliability and subjectivity of judges are crucial since even expert judges can differ in creating clone reference data [50]. Handling different clone types, notably "near-miss" clones with statement-level variations, poses another challenge. As Bellon's benchmark, though valuable, requires a reference corpus update, it underscores the ongoing evolution of clone detection assessments beyond its scope [3]. In our benchmark, we have multiple judges to validate the data, and we tested different clone detection tools with our benchmark to justify that our benchmark consists of only semantic clones. Apart from that, GPTCloneBench has clone pairs of four different programming languages.

Similarly, Krutz and Le [17] chose 1536 method pairs randomly from three open-source C programs: PostgreSQL, Python, and Apache. Four students and three experienced judges were hired to manually assess these couples. Out of the 1536 candidate clones, they discovered that only 66 clone pairings were true clones. Their benchmark includes 9 type-4 (semantic clones), 14 type-3, and 43 type-2 clones. While their benchmark has high confidence, it lacks the size and variety needed to reliable measure clone detection recall. Compared to that, we have a huge number of semantic clone pairs along with support for Java, Python, C and C# languages.

Roy and Cordy [32] developed a number of situations to produce diverse clones, which are then injected into the code base and used in the assessment process to gauge recall and precision. However, these scenarios must be thorough (covering all varieties of clones that might be present in actual source code) and independent of any clone definition. These clones have been created using a mutation-based approach and do not really represent real clones.

Recently, Yuki et al. [55] developed a method to create a benchmark by identifying merged methods (merged cloned methods in the next version) by mining software versions. If two merged methods are called by the same methods in the following revision and share a textual similarity, they are regarded as clones. However, their method is restricted to refactored clones, which comprise a very small percentage of code clones. Only 19 clones out of more than 15K variants could be found.

Svajlenko et al. [46] [42] have introduced BigCloneBench, which is mined from a large inter-project source repository by targeting the implementation of 43 functionalities. BigCloneBench is built from the IJaDataset 2.0, a dataset of 250M LOC in 2.5M Java files from 25K projects mined from SourceForge and Google Code. It has 48 thousand Type-1 clones, 4.2 thousand Type-2 clones, and 8.9 million Type-3/Type-4 clones. The split calculation for Type-3 and Type-4 by their measured syntactical similarity is 34 thousand 70-90% similarities, 329 thousand with 50-70% similarity, and 8.5 million with 0-50% similarity [46] [42]. However, while BigCloneBench is best suited for comparing and evaluating modern clone detection tools (up to Type-3 detectors), it are not optimal for evaluating semantic clone detection tools [47]. Furthermore, while BigCloneBench has been widely used for machine learning based semantic clone detection, there are a number of limitations identified recently [16]. Also, it has support for Java programming language only. In contrast to this, in our benchmark, we have more semantic clones with an almost equal number of semantic and non-semantic clone pairs, along with quality semantic clones. This makes the benchmark suitable for ML training. We have also introduced more language support compared to BigCloneBench.

To improve the semantic behaviour of the BigCloneBench dataset, Yu et al. [54] proposed an updated version of the dataset focused on by abstracting the identifier names. The authors of [54], however, just updated the identifier names, ignoring any potential implementation changes that might have been made while maintaining the same functionality. As a result, it is not clear whether they would have got a similar output with clone pairs, which were implemented in different ways i.e., changing the logic but giving the same functionality.

Al-omari et al. [1] built SemanticCloneBench, a dataset of semantically equivalent code snippets intended to help researchers compare and evaluate techniques for detecting semantic clones. Even though it has semantic clone pairs of four different languages, unfortunately, the number of clone pairs is very low to train an ML model. In addition, we have found some syntactical issues in the benchmark, along with a mixture of C++ code in the C programming language. Compared to this, our benchmark is 14-fold larger, consisting of semantic and cross-language clone pairs with no syntactic error.

Nafi et al. [26] have introduced a new clone pair benchmark for cross-language named CLCDSA. CLCDSA utilizes syntactic features and API documentation for cross-language clone detection. Nafi et al. evaluated CLCDSA on a dataset of code snippets in Java, Python, and C++. Unfortunately, this dataset is considered toy data because of the lack of real-world examples. This benchmark is generated from code examples (the solutions of a given problem for example) of programming competitions, which has little to no use in real-world systems. Even though our benchmark was developed through Generative AI, the input code fragments are close to real-world examples as they have been taken from StackOverflow. Besides, we have introduced C programming language support along with a more functional approach.

## VII. THREATS TO VALIDITY

The major concern for our benchmark is that we have used generative AI technology to develop the benchmark. As a result, these may not be real clones. It is evident that clones can be real-world or artificial clones [33]. If a human writes a code fragment, that is called a real clone; otherwise artificial or generated clone. We understand that our clones are kind of artificial clones. However, we have used SemanticCloneBench to generate the results. SemanticCloneBench clones are real because they are created based on provided knowledge by developers who participate in the crowd-sourced information website Stack Overflow. Because of that, the GPT-generated code fragments are a similar kind of code that refers to real code because of how we used SemanticCloneBench data as our input. We are not saying the clones generated by GPT are real world, but because of how we formulated the query, we can say it is in between real-world and artificial clones. We tested by compiling both the generated code and the given input code and received the same results. We also analysed the GPT-generated code and found the code is more well-structured, functional and object-oriented, which can be helpful for programmers.

In order to avoid any bias in the manual validation process, we hired judges who have the necessary knowledge and experience in software engineering and system development. For the decision conflict, as we mentioned earlier, we used three post-docs to resolve any conflict through discussion. As a result, the decision mostly remains unbiased. Still, we agree that manual validation can introduce some errors, and we are trying to add more judges in future or make the system more automated to reduce the human error percentage.

As GPT-3 generate different code (output) for a given input and prompt, that is why we followed a formal prompt engineering method [53] [27] [20] while making up the queries and tested many other queries other than those mentioned in our paper. We only selected the prompts with best results for our work. So even if someone creates a new prompt, we believe they will get similar outcomes to our results and could confirm our findings.

Finally, the biggest question is, "Are we trying to create a benchmark for training machine learning-based clone detection tools by other machine learning-based techniques like GPT-3? and how feasible is it?". Although we are using a language model to create the semantic clones for us, the output has to go through a lot of processing and filtering to get listed into the benchmark. So, even though the code fragments are machine-generated, the benchmark has all the filtered and processed clone pairs that can be used in machine learning-based clone detection tools because in our prompt engineering to create these clones, the main input is taken from stack overflow, which is a real-world software developer's community. As a result, the GPT-3 followed the real-world code to generate new code. Additionally, generated clones by GPT-3 are potentially human-like code because of the nature of how the GPT model was trained with a large corpus of human-generated real-world codes, and it has learned to recognize patterns and structures in code that allows it to generate new human-like code [7]. Furthermore, this benchmark has other applications than using as training dataset for building machine learning based clone detection tools. This benchmark can now be used to evaluate and compare the semantic and cross-language clone detection tools including those detectors that detect clones across Microsoft .NET programming languages [1]. This benchmark could help cross check whether source transformation based clone detection tools such as CloneWorks [45] or SimCad [49] could in fact detect semantic clones, could help build IDE-based flexible clone detection and management [56], [57], or even could potentially be used in building similar benchmarks in other contexts [30]. One could thus safely conclude that even though GPTCloneBench is mostly machine generated, it has potential in contributing to Software Engineering and beyond.

## VIII. CONCLUSION

Our research proposes a novel approach for building a comprehensive benchmark of semantic and cross-language clones using the GPT-3 model. We attempted to address the limitations of existing benchmarks and their difficulties in recognising semantic and cross-language clones. We removed clone pairs with textual similarity and used nine judges to validate our benchmark to improve outcomes. To prove our clone pairs are semantic clones, we evaluated multiple clone detection tools using our benchmark. Our Benchmark has more semantic clone pairs than SemanticCloneBench, more programming language support than BigCloneBench with the resolved imbalance and labelling problems for semantic clones, and more functional, additional programming language implementation, and reusable code than the CLCDSA dataset, allowing the software clone community to do more research on semantic clones. This initial edition of the GPTCloneBench includes over 37,149 true semantic clones, 20,770 cross-language pairs and 19,288 false semantic(Type-1/Type-2) clone pairs. In the next version, we will add more data to our benchmark with the unprocessed data that we have for other different prompts and some false clones.

## ACKNOWLEDGMENT

This work was supported in-part by the Natural Sciences and Engineering Research Council of Canada (NSERC) Discovery grants, the John R. Evans Leaders Fund (JELF) of the Canada Foundation for Innovation (CFI), and NSERC CREATE graduate program on Software Analytics Research (SOAR) grants.